\documentclass[10pt]{article}

\usepackage{cite} %to have multiple citations

\usepackage{graphicx}
\usepackage[font=scriptsize]{caption}

\usepackage[a4paper,top=2cm,bottom=2cm,left=3cm,right=3cm]{geometry} %per allargare i margini

\usepackage{hyperref}

\begin{document}

\title{\textbf{School visits to a physics research laboratory using
virtual reality}\vspace{-4ex}}
\date{}
\maketitle
\begin{center}
\author{Ilaria De Angelis$^{1,2}$, Antonio Budano$^{2}$, Giacomo De Pietro$^{2}$,\\ Alberto Martini$^{3}$ and Adriana Postiglione$^{1,2}$}
\end{center}

\paragraph{} \parbox[t]{1\columnwidth}{$^1$Dipartimento di Matematica e Fisica, Universit\`a degli Studi Roma \\Tre, Rome (Italy)\\%
    $^2$INFN Sezione di Roma Tre, Rome (Italy)\\
    $^3$Deutsches Elektronen–Synchrotron, 22607 Hamburg (Germany)\\
    
    ilaria.deangelis@uniroma3.it}

\begin{abstract}
School visits to research laboratories or facilities represent a unique way to bring students closer to science and STEM (Science, Technology, Engineering and Mathematics) careers. However, such visits can be very expensive for students and teachers, in terms of both time and money. In this paper, we present a possible alternative to on-site visits consisting in an activity addressed to high school students that makes use of a VR application to make them “enter” into a particle physics experiment. This proposal can represent a valid way of guaranteeing a visit to a research centre for all schools, regardless of their social or geographical origin. We describe the tests we carried out with a focus group of teachers and their students, and the obtained results. 
\end{abstract}


\section{Introduction}
Guided visits to research centres or facilities certainly represent a peculiar element in a student's high school career, since they allow direct contact with authentic conditions of scientific knowledge production processes \cite{Dimopoulos}. In the Italian National Indication guidelines on teaching \cite{Indication}, in fact, these visits are explicitly mentioned for physics, as they represent one of the means by which students reach their learning objectives at the end of their high school career. Experiencing some time in a research centre can indeed not only improve students’ knowledge of physics, but also lead to a clearer idea of what research in physics is about and eventually motivate some of them to consider a science profession \cite{Neresini}. Therefore, these visits should become part of a scientific school curriculum, along with hands-on and practical activities \cite{Snetinova,Sokolowska, Postiglione1, Postiglione2, Postiglione3}. 

In Italy, two examples of internationally renowned laboratories that organise visits addressed to school groups are the Laboratori Nazionali del Gran Sasso \cite{lngs} and the Sardinia Radio Telescope \cite{srt}. In Europe, CERN is one of the most active centres as regards proposals for schools \cite{Ellis}. Worldwide, several research centres or facilities open their doors to schools.  
The participation of students and teachers to visits, however, although certainly meaningful, can be very expensive in terms of money and time, especially if the centres are located in places far from the school. For this reason, physics teachers often choose alternative activities to ensure contact with research organisations that do not require an on-site visit. An example in this sense are the CERN International Masterclasses \cite{Cecire1, Cecire2}, which allow students to work from their schools on real particle physics data, and discuss the related analysis together with CERN researchers during a video-conference. In this way, participants can virtually walk into a scientific central control room and get a glance of what they would see entering CERN. 
The advantages of initiatives of this kind are manifold, from becoming aware of the frontiers of scientific research, to actively working on real data, to coming into contact with
an international research environment \cite{DeAngelis}. On the other hand, however, the contact with the laboratory is only provided by the video-conferences that generally involve many students’ groups at the same time \cite{Cecire1}.
In this context, we worked to develop a third way, alternative to both face-to-face visits and masterclass-type initiatives, through which a student can experience the world of a scientific research laboratory up close. Our approach makes use of Virtual Reality (VR) technology. To do this, we chose the context of particle physics, in particular the Belle II collaboration, for which an advanced VR application was developed \cite{DeAngelis}.  
The remaining paper is organised as follows. In section 2 we describe the activity we developed. In section 3 we illustrate the public we reached, including both students and teachers, and the feedback we received and in section 4 we present our conclusion.

\section{The educational proposal}
We have chosen to organise a virtual visit to an international laboratory which is however very difficult and expensive to reach for Italian school groups, since it is located in Japan (much further away than Laboratori Nazionali del Gran Sasso or CERN). In fact, our educational proposal for schools initiated from the VR application Belle2VR \cite{Duer}.
Having realised the potential of Belle2VR application, we soon started to use it with the public during some outreach events such as science festivals or public events organised at the University. In these cases, visitors were given the possibility to wear the VR helmet while the researchers used joysticks to guide them to discover the experiment, as in a real guided visit.
After a few years of experience in science festivals and open events to which thousands of people participated, including many school students, that provided us very positive feedback, we have decided to propose a more structured activity to schools.

\subsection{Belle2VR}
Developed by Virginia Tech, the application Belle2VR allows users to virtually enter the particle physics detector of the Belle II experiment \cite{Kou}. The Belle II experiment is currently carried out at the KEK in Tsukuba, Japan, and it studies the properties of heavy quarks and leptons to search for an evidence of new physics phenomena, from the matter-antimatter asymmetry problem to the existence of dark matter particles. Belle2VR reconstructs the interior of the detector and allows to visualise realistic simulations of particles interacting with each other and with the detector elements (Fig. 1). The user can navigate through the detector and its components and can also manage the time evolution of the interaction by going back and forth or stopping the Developed by Virginia Tech, the application Belle2VR allows users to virtually enter the particle physics detector of the Belle II experiment \cite{Kou}. The Belle II experiment is currently carried out at the KEK in Tsukuba, Japan, and it studies the properties of heavy quarks and leptons to search for an evidence of new physics phenomena, from the matter-antimatter asymmetry problem to the existence of dark matter particles. Belle2VR reconstructs the interior of the detector and allows to visualise realistic simulations of particles interacting with each other and with the detector elements (Fig. 1). The user can navigate through the detector and its components and can also manage the time evolution of the interaction by going back and forth or stopping the motion of particles at a specific time. Belle2VR, therefore, allows to explore particle physics phenomena from a unique point of view.

\subsection{Activity structure}
We built an activity addressed to high school class groups, lasting about an hour and a half, that can be carried out in a dedicated University room, or directly in the classroom. 
It starts with a theoretical introduction that makes use of slides. Here, some basic topics and concepts typically treated at school are recalled, such as electromagnetism. At the same time, more recent contents are also presented, such as the Standard Model, the cross section or the decay of particles, which require the use of quantum physics. The Belle II experiment is also presented in terms of its components and physics goal.

\begin{figure}[ht]
  \centering
\includegraphics[width=0.5\columnwidth]{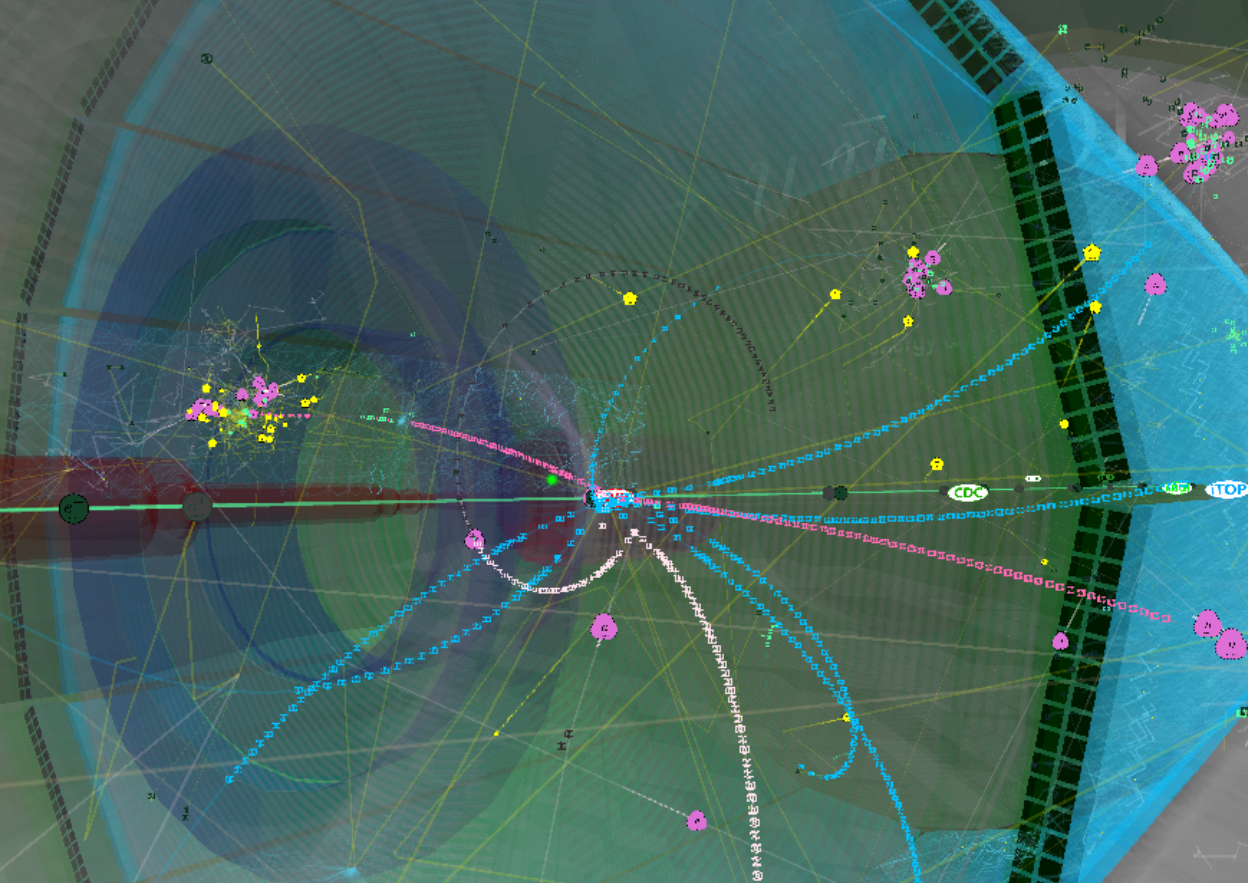}
\caption{Snapshot of a simulated event into the Belle2VR application.}
\end{figure}

This phase is meant to represent the welcome and introduction step that characterises the initial part of a typical on-site visit to a research laboratory \cite{Neresini}.
Subsequently, the researcher puts on the VR helmet while a large screen shows to the group what he/she sees. At that point, participants enter the detector for the first time together with the researcher. He/she moves in the virtual environment by movements of the head, allowing to display the detector details and some collisions between particles that have been selected by he/she. This allows to underline the most important aspects of the experiment and to visualise what researchers described in the first part of the activity. This is the moment in which students access the researcher's work environment, and begin to look at it through his/her eyes and his/her emotion. 
At this point students in turn put on the helmets, enter the detector in first person and explore the virtual space while a researcher stays close to him/her to guide him/her and answer all his/her questions and curiosities. Usually, we dedicate from two to three researchers in the activity, so that we can carry on this phase using up to three VR parallel stations.

In the meantime, the rest of the group watches their classmate while living the experience and follows the discussion with the researcher.

\section{Collection of data and results}
Once the activity design was completed, we tested it with students of different ages and schools. To do this, we first involved some of the teachers already used to work with us in testing, discussing and optimising innovative activities. Together with them, we selected 7 groups of students (one for each teacher) from different schools: 2 classes of the fifth and final year of high school (17-18 years old), 2 classes of the fourth year (16-17 years old), 1 class of the third year (15-16 years old) and 2 mixed groups of third, fourth and fifth year students. In this way, we had both groups of students all very interested in learning more about physics (the two mixed groups) and typical school classes where interested and non-interested students coexist. Regarding the school type, the vast majority of participants attended the “Liceo Scientifico”, i.e. the Italian high school focused on science subjects; only one mixed group of students attended the “Liceo Classico”, the Italian high school focused on the humanities.
After carrying out the activity with the students in the presence of their teachers, we asked the latter to talk with their class to get their impressions on our proposal. Later, we carried out open interviews with all participating teachers separately.

\subsection{Results}
In general, the activity was very positively received by both teachers and students. In fact, 5 out of 7 teachers told us that their students voted 5 out of 5 and 2 out of 7 teachers told us their students voted 4 out of 5 to the activity from a general point of view. The teachers' score was also very positive, as 6 out of 7 teachers voted 5 out of 5 and 1 teacher  voted 4 out of 5.
At this point, we asked for more details on their vote. Specifically, we first asked them what they particularly liked about the activity. Three of them told us that they enjoyed the use of VR technology; one teacher stated that the strength of the activity lays in the possibility of getting inside the particle detector; another teacher appreciated the opportunity of “directly seeing” what it means doing research with a particle accelerator; one teacher mentioned the possibility of bringing the world of research closer to students; another teacher especially appreciated the clarity of the researchers who carried out the activity. 
Then, we asked their opinion about the different phases of the activity. The introductory part, realised using slides, was considered clear and well organised by all the teachers. Two teachers also pointed out that some topics could be deepened, such as the concept of interaction between particles and the mass-energy equivalence. The part of the activity that makes use of Belle2VR has been defined by all teachers as interesting, fun and engaging.
As for the negative aspects of the activity, the majority of the teachers stated that they couldn’t find any; the only elements raised by two teachers concerned the limited number of students that can be involved and the role of some participants considered too passive. 

Subsequently, we asked the teachers what objectives they think the activity was able to achieve. Some answers concerned the possibility of understanding and visualising particle physics (one teacher in particular stated that his students even understood the uncertainty principle thanks to the activity). Other answers cited the possibility of inspiring curiosity and interest toward physics and science, and of bringing students closer to the work of a physicist.
At the end of the interview we explicitly asked the teachers which class year is more suitable for the activity and if they would have proposed the activity to other classes. The majority stated that the activity is suitable for the final months of the fourth year or the fifth year (when Italian students have typically already dealt with electromagnetism and a first introduction of quantum physics). Two teachers, however, claimed that even third-year students can benefit from the activity, as it is fascinating and inspiring. All the teachers claimed that they would surely recommend the activity to other classes.

\section{Discussion and conclusion}
In this paper we presented an educational proposal addressed to high schools and realised at our University that makes use of the VR technology to enter a physics research laboratory. The activity aims to constitute an alternative proposal to on-site visits to research centres, which, while particularly formative and enriching for students, are also very expensive in terms of time and money. Our proposal retraces all the stages of an on-site visit \cite{Neresini}: welcoming and introduction; entering into the laboratory or facility; interaction and discussion with the public. Throughout the initiative, a fundamental role is played by the researchers who carry out the activity. In fact, they not only guide the public in the laboratory (in our case piloting the Belle2VR application) but also share their emotions and experiences with students, thus helping to paint a realistic representation of their working environment. 
Following the discussion with a focus group of 7 high school teachers who participated in the activity together with their classes, we can state that our proposal was very well received by school and therefore we are strongly motivated to replicate it with other school groups in the future.
In fact, the teachers greatly appreciated the activity. They underlined several aspects that this proposal manages to achieve: visualising and understanding phenomena otherwise impossible to see such as those related to particle physics; spreading VR technology; intriguing students about physics and science; giving participants a more realistic view of the scientific research world and of the work of a scientist. All these elements contribute to strengthening physics teaching and bringing students closer to STEM careers.
The teachers also helped us to identify some aspects we can work on to improve our activity: the limited number of students that can be involved and the too passive role experienced by a small part of them. These aspects seem to be easily overcome, for example adding more parallel VR stations, where more students can virtually enter the experiment at the same time.

A very significant aspect of our proposal consists in the possibility of involving schools easily in any place without them having to face high travel expenses or heavy time commitment. In this sense, our initiative could provide a valuable example of a method to introduce a visit to a research laboratory on a permanent basis in physics school curricula of all students, regardless of their availability of financial resources and their geographical location.
For this reason, we believe that our proposal is worth being exported to other research centres or facilities, even in fields other than particle physics.

\section*{Acknowledgements}
This work was supported by the Italian Project ‘Piano Lauree Scientifiche’. We thank the teachers and students who participated in our activity.


\begin{thebibliography}{16}

\bibitem{Dimopoulos}
Dimopoulos K, Koulaidis V, Int. J. of Learn. Ann. Rev. 12 (2006) 10\\ \url{http://dx.doi.org/10.18848/1447-9494/CGP/v12i10/48219}

\bibitem{Indication}
Italian National Indication, Ministry of Education, 2010\\
\url{https://www.istruzione.it/alternanza/allegati/NORMATIVA\%20ASL/INDICAZIONI\%20NAZIONALI\%20PER\%20I\%20LICEI.pdf}

\bibitem{Neresini}
Neresini F, Dimopoulos K, Kallfass M and Peters H P, Sci. Comm. 30 (2009) 506\\ \url{https://doi.org/10.1177\%2F1075547009332650}

\bibitem{Snetinova}
Sn\v{e}tinov\'a M and K\'acovsk\'y P 2019 J. Phys.: Conf. Ser. 1287 012049 \\
\url{https://doi.org/10.1088/1742-6596/1287/1/012049}

\bibitem{Sokolowska}
Soko\l owska D and Michelini M The Role of Laboratory Work in Improving Physics Teaching and Learning (2018) Springer Cham \\
\url{https://doi.org/10.1007/978-3-319-96184-2}

\bibitem{Postiglione1}
Postiglione A and De Angelis I Phys. Educ. 56 (2021) 025019 \\
\url{https://doi.org/10.1088/1361-6552/abcab4}

\bibitem{Postiglione2}
Postiglione A and De Angelis I, Phys. Educ. 56 (2021) 025020 \\
\url{https://doi.org/10.1088/1361-6552/abd1c4}

\bibitem{Postiglione3}
Postiglione A,  Il Nuovo Cimento 45 C(2022) 91 \\
\url{http://dx.doi.org/10.1393/ncc/i2022-22091-x}

\bibitem{lngs}
\url{https://www.lngs.infn.it/en/educational}

\bibitem{srt}
\url{http://www.srt.inaf.it/outreach/guided-tours-srt/}

\bibitem{Ellis}
Ellis J (2000) \url{https://doi.org/10.48550/arXiv.physics/0005021}

\bibitem{Cecire1}
Cecire K. (2011) DPF-2011 Conference \\
\url{https://doi.org/10.48550/arXiv.1109.2559}

\bibitem{Cecire2}
Cecire K and Dower R, DPF2019 Conference(2019) \\ \url{https://doi.org/10.48550/arXiv.1910.00522}

\bibitem{DeAngelis}
De Angelis I, Postiglione A, La Franca F,  Il Nuovo Cimento C 4-5 (2021) 162 \\ \url{http://dx.doi.org/10.1393/ncc/i2021-21162-x}

\bibitem{Duer}
Duer Z, Piilonen L and Glasson G, IEEE Comp. Graph. and App. 38 (2018) 3 33 \\ 
\url{https://doi.org/10.1109/MCG.2018.032421652}

\bibitem{Kou}
Kou et al., Prog. Theor. Exp. Phys., 12 (2019) 123C01, 2019 \\ 
\url{https://doi.org/10.1093/ptep/ptz106}

\end{thebibliography}
\end{document}